\documentclass{article}
\usepackage[OT1]{fontenc} 
\usepackage{spconf,amsmath,amssymb,amsfonts,graphicx}
\usepackage{cite}
\usepackage{mathrsfs}
\usepackage{algorithm}
\usepackage{algorithmic}
\usepackage{enumerate}
\usepackage{url}
\usepackage{multirow}
\usepackage{fancyhdr}
\usepackage{color}

\title{Phase-aware harmonic/percussive source separation \\ via convex optimization}
\name{Yoshiki Masuyama, Kohei Yatabe and Yasuhiro Oikawa \vspace{-4pt}}
\address{\fontsize{11pt}{0pt}\selectfont Department of Intermedia Art and Science, Waseda University, Tokyo, Japan}
\begin{document}
\ninept
\maketitle
\begin{abstract}
\vspace{-1pt}
Decomposition of an audio mixture into harmonic and percussive components, namely harmonic/percussive source separation (HPSS), is a useful pre-processing tool for many audio applications.
Popular approaches to HPSS exploit the distinctive source-specific structures of power spectrograms.
However, such approaches consider only power spectrograms, and the phase remains intact for resynthesizing the separated signals.
In this paper, we propose a phase-aware HPSS method based on the structure of the phase of harmonic components.
It is formulated as a convex optimization problem in the time domain, which enables the simultaneous treatment of both amplitude and phase.
The numerical experiment validates the effectiveness of the proposed method.
\vspace{-1pt}
\end{abstract}

%
\vspace{-4pt}
\section{Introduction}
\vspace{-4pt}

Audio source separation, decomposing an audio mixture into each source, is one of the fundamental tools for audio signal processing.
In particular, harmonic/percussive source separation (HPSS), which decomposes an audio mixture into harmonic components (e.g., guitar and piano) and percussive components (e.g., drums), has gained much attention as a pre-processing tool for many music-information retrieval (MIR) tasks including chord estimation \cite{HPSSApp1} and tempo estimation \cite{HPSSApp2}.
For instance, extracted percussive components are useful cues for tempo estimation while the harmonic components are crucial for chord estimation.
HPSS is also helpful for audio remixing \cite{HPSSApp3} and time-scale modification \cite{HPSSApp4}.

\begin{figure}[t]
	\centering
	\includegraphics[width=0.95\columnwidth]{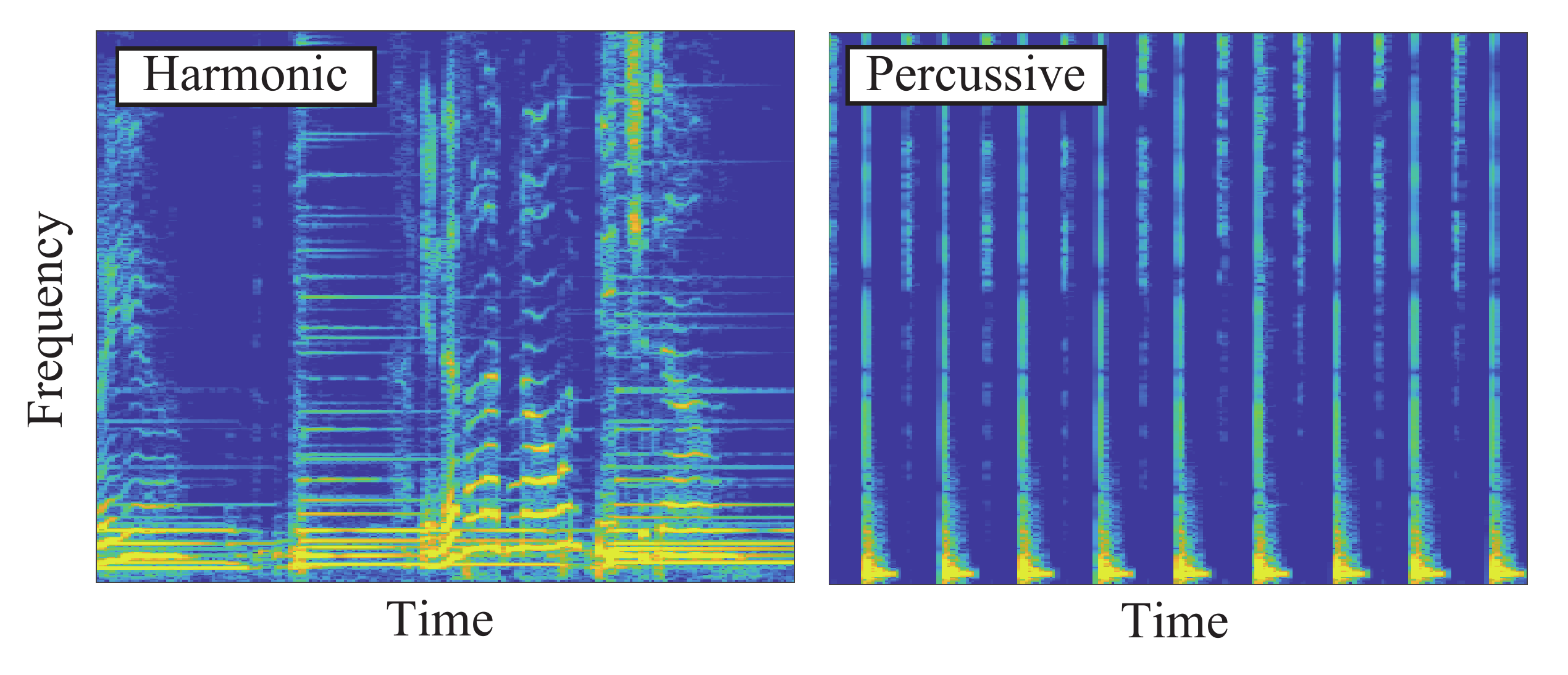}
	\vspace{-12pt}
	\caption{Example of spectrograms of the harmonic and percussive components. The harmonic component is continuous in the time direction (left), and that of percussive components is continuous in the frequency direction (right). The time-frame-wise sparse structure is apparent in that of the percussive component.}
	\label{fig: HP}
\vspace{-8pt}
\end{figure}

One of the main approaches to HPSS is to take advantage of the \textit{anisotropic smoothness} of power spectrograms (i.e., power spectrograms of harmonic components are continuous in the time direction, and those of percussive components are continuous in the frequency direction as shown in Fig.~\ref{fig: HP}) \cite{HPSS2008, HPSS2012, HPSS2014Ono}.
Based on the anisotropic smoothness, HPSS was formulated as an optimization problem of minimizing the $\ell_2$ norm of the gradient of power spectrograms in \cite{HPSS2008}.
Considering the same assumption, the method presented in \cite{HPSS2010} applies the time-/frequency-directional median filtering (MF) to the power spectrogram of the audio mixture, which was further developed into the kernel additive model (KAM) \cite{KAMOri, HPSS2014KAM, HPSS2018ICASSP}.
Although these methods have been successfully applied to HPSS, they have a limitation because the degraded phase from the audio mixture is still utilized for resynthesizing the separated time domain signals.

Recent literature has shown the importance of phase in audio source separation \cite{NMFBench, PU2} and audio denoising \cite{PCTV, iPCTV, IWAENC2018, iPCLR}.
These studies utilize a model of phase for harmonic components, called \textit{sinusoidal model}.
This model claims that the phase evolution of harmonic components can be predicted from their instantaneous frequencies.
More recently, this model was also applied to HPSS \cite{HPSS2014Phase, HPSS2018IWAENC}.
In \cite{HPSS2014Phase}, the real-valued time-frequency mask for extracting harmonic components was constructed based on the sinusoidal model.
However, the phase is not modified through time-frequency masking, and thus the degraded phase from the audio mixture is utilized for resynthesizing back to the time-domain.
On the other hand, \cite{HPSS2018IWAENC} utilizes the sinusoidal model for recovering the phase after applying a time-frequency mask obtained by a deep neural network.
However, simultaneous modification of both amplitude and phase has not been presented for HPSS.

In this paper, we propose a phase-aware HPSS method, which treats both amplitude and phase simultaneously, through convex optimization in the time domain.
For harmonic components, the proposed method assumes the time-directional smoothness of the complex-valued spectrogram with the recently proposed phase modification \cite{iPCTV}.
This assumption can be interpreted as the unification of the two conventional HPSS methodologies: the anisotropic smoothness and sinusoidal model.
For the percussive components, the time-frame-wise sparsity is assumed on their complex-valued spectrograms, which does not require any assumptions for its phase structure.
The effectiveness of the proposed method was confirmed by the signal-to-distortion ratio (SDR).

\vspace{-4pt}
\section{Previous works}
\vspace{-2pt}

In this section, we briefly revisit two approaches to HPSS, the anisotropic smoothness and sinusoidal model, since the proposed method combines these two approaches as described in Section~\ref{sec: proposed}.

\vspace{-2pt}
\subsection{HPSS based on anisotropic smoothness}
\label{sec: ono}
\vspace{-2pt}

One main approach to HPSS is the anisotropic smoothness of power spectrograms.
It assumes that a power spectrogram of harmonic components $\mathbf{H} \in \mathbb{R}_+^{K \times T}$ and that of percussive components $\mathbf{P} \in \mathbb{R}_+^{K \times T}$ have the following relations:
\begin{align}
H_{\omega, \tau} &\approx H_{\omega, \tau \pm1},
\label{eq: hsmooth}\\
P_{\omega, \tau} &\approx P_{\omega \pm1, \tau},
\label{eq: psmooth}
\end{align}
where $\omega  = 1, \ldots, K$ and $\tau  = 1, \ldots, T$ are frequency and time indices, respectively.
Eq.~\eqref{eq: hsmooth} indicates that the power spectrogram of the harmonic components varies slowly, while Eq.~\eqref{eq: psmooth} claims that of the percussive components is smooth in the frequency direction.
Based on these assumptions, the following optimization-based HPSS method was proposed in \cite{HPSS2008}:
\begin{equation}
\begin{array}{cl}
    \displaystyle\min_{\mathbf{H}, \mathbf{P}} &\! \displaystyle\frac{1}{2\sigma_\mathrm{h}^2} \left\| D_\tau( \mathbf{H} )\right\|_\text{Fro}^2
+ \frac{1}{2\sigma_\mathrm{p}^2} \left\| D_\omega( \mathbf{P} )\right\|_\text{Fro}^2 \\[10pt]
\text{s.t.} &\! \displaystyle H_{\omega, \tau} +  P_{\omega, \tau} =  | X_{\omega, \tau}|^{2\gamma}, \quad H_{\omega, \tau}\geq0, \quad P_{\omega, \tau}\geq0
\end{array}\!,\!
\label{eq: ono}
\end{equation}
where $D_\tau$ and $D_\omega$ are the time and frequency-directional differences (i.e., discrete approximation of directional derivatives), $\sigma_\mathrm{h}$ and $\sigma_\mathrm{p}$ are parameters to adjust smoothness of harmonic and percussive spectrograms, and $\mathbf{X}\in\mathbb{C}^{K\times T}$ is the complex-valued spectrogram of the audio mixture to be separated.
$\gamma$ is a hyperparameter for range compression $(0\!<\!\gamma\!\leq\!1)$, and it will be set to $1$ in the rest of this paper for simplicity.
By minimizing the energy of directional derivatives of the spectrograms, this model attempts to find spectrograms which are smooth in each direction.
In the experimental section, this optimization-based method will be referred to as \textit{Ono's}.

While this optimization-based method is simple and effective, the assumption of additivity of power spectrograms holds only approximately (it can be justified only by some statistical sense \cite{ISNMF}).
Furthermore, it considers only power spectrograms, and thus the phase information is ignored.
Although some extensions based on the anisotropic smoothness of power spectrograms have been proposed \cite{HPSS2010, KAMOri, HPSS2014KAM, HPSS2018ICASSP}, the degraded phase of the audio mixture is utilized for resynthesizing the separated time domain signals, which often causes audible artifacts as mentioned in \cite{NMFBench}.

\vspace{-2pt}
\subsection{HPSS based on sinusoidal model}
\vspace{-2pt}
Another recent approach to HPSS is based on the sinusoidal model.
Let the short-time Fourier transform (STFT) of a signal $\mathbf{x} \in \mathbb{R}^{L}$ be
\begin{equation}
\mathscr{F} (\mathbf{x})_{\omega, \tau} = \sum_{l = 0}^{L-1} {x}_{l+a\tau} \, g_l \, \mathrm{e}^{-2 \pi j \omega b l / L},
\label{eq: STFT}
\end{equation}
where $\boldsymbol{g} \in \mathbb{R}^{L}$ is a window, $j = \sqrt{-1}$, $a$ and $b$ are the time and frequency shifting steps, and index overflow is treated by zero-padding.
Considering a sinusoid given by
\begin{equation}
s_l = {A} \, \mathrm{e}^{2 \pi j f l/L + \phi},
\label{eq: sinusoid}
\end{equation}
where ${A} \in \mathbb{R}_+$, $f \in [0, L/2)$, and $\phi \in [0, 2\pi)$ are the amplitude, frequency, and initial phase, respectively.
Its phase spectrogram $\boldsymbol{\phi}$ (with appropriate unwrapping) has the following relation: 
\begin{equation}
\phi_{\omega, \tau} = \phi_{\omega, \tau-1} + 2 \pi a v_{\omega, \tau-1},
\label{eq: sinusoidal model}
\end{equation}
where $v_{\omega, \tau}$ is the instantaneous frequency at each time-frequency bin.
This sinusoidal model has been studied in phase vocoders \cite{PV}, and applied to audio signal processing tasks recently \cite{PU1, PU2, IWAENC2018, iPCLR}.

More recently, the sinusoidal model was also applied to HPSS \cite{HPSS2014Phase, HPSS2018IWAENC}.
The phase-based masking (PM), which constructs the time-frequency mask based on the relation among phases in successive time frames, was presented in \cite{HPSS2014Phase}.
Although this method considers phase information, the phase of the audio mixture is still utilized for resynthesizing the separated time domain signals.
In contrast, the method presented in \cite{HPSS2018IWAENC} utilizes the sinusoidal model for modifying phase.
Specifically, the time-frequency mask is estimated by a deep neural network, and phases of separated signals are estimated by the specific algorithm based on the sinusoidal model \cite{PU2}.
It significantly depends on the time-frequency mask estimation, and phase information is utilized just in the post-processing.

\vspace{-2pt}
\section{Proposed HPSS method}
\label{sec: proposed}
\vspace{-2pt}

In this section, we propose a phase-aware HPSS method through convex optimization where the phase-aware smoothness in the time direction is assumed for harmonic components as a unification of the aforementioned approaches: anisotropic smoothness and sinusoidal model.
On the other hand, the time-frame-wise sparsity is considered for the complex-valued spectrograms of percussive components as a phase insensitive prior.
The proposed method directly separates the time-domain signal, which enables the simultaneous modification of amplitude and phase.

At first, we review the phase-aware smoothness of harmonic signals introduced in the previous study \cite{iPCTV}.
Then, based upon that, we formulate the proposed method and discuss the relation to the conventional HPSS approaches reviewed in the previous section.

\begin{figure}[t]
	\centering
	\includegraphics[width=0.96\columnwidth]{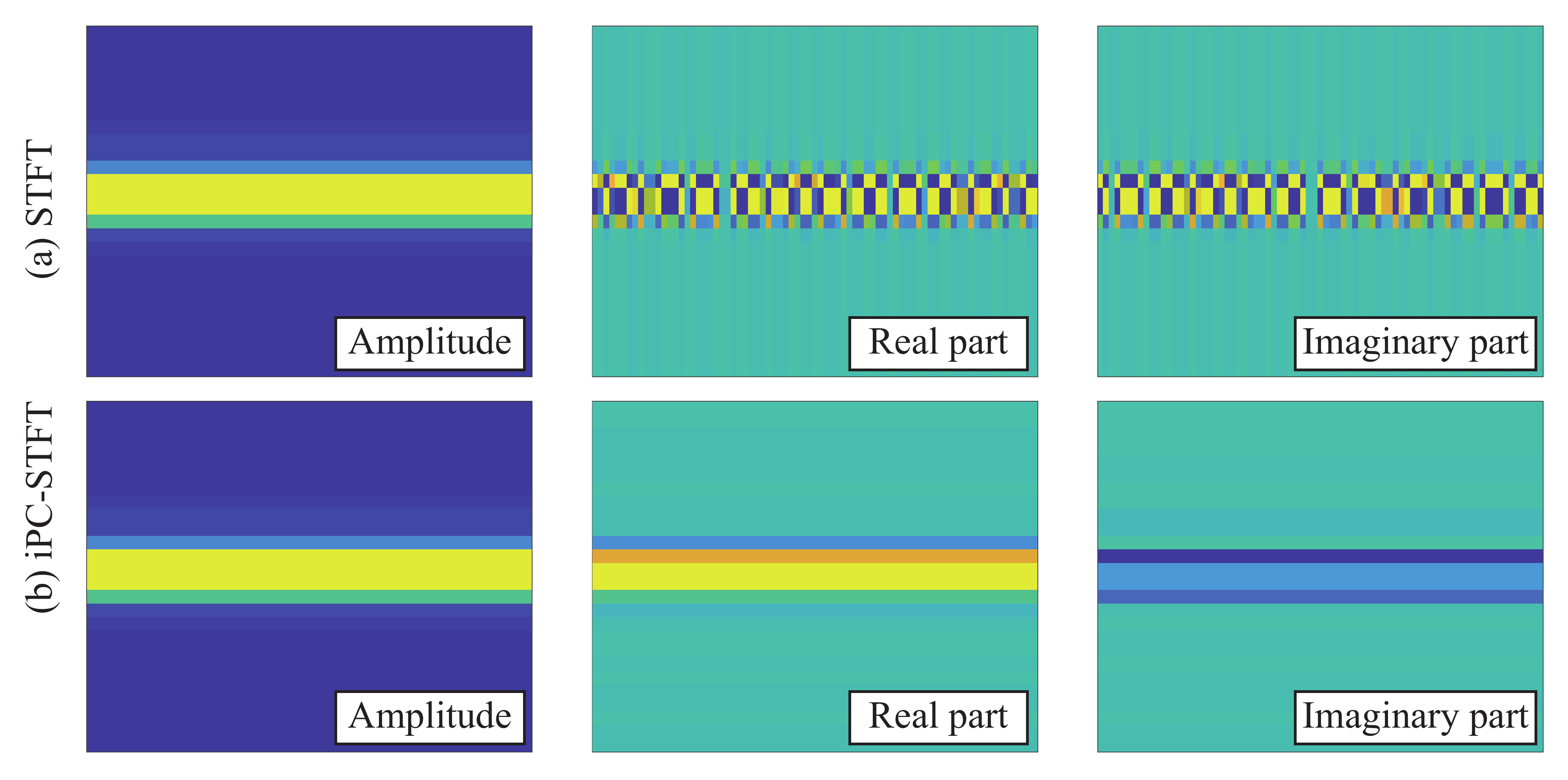}
	\vspace{-8pt}
	\caption{Illustration of a complex-valued spectrogram calculated by (a) the usual STFT and (b) iPC-STFT.}
	\label{fig: illust}
\vspace{-8pt}
\end{figure}

\vspace{-2pt}
\subsection{Time-directional smoothness of instantaneous phase corrected complex-valued spectrogram}
\label{sec: iPCTV}
\vspace{-2pt}

As shown in Eq.~\eqref{eq: sinusoidal model}, the phase of a sinusoid has the relation among successive time frames.
It indicates the phase in the next frame is predictable from the phase and instantaneous frequency in the current frame.
Based on the relation of phase, a model of a complex-valued spectrogram of harmonic components was recently studied for audio signal processing \cite{PCTV, iPCTV}.
Since the amplitude of the sinusoid is constant, its complex-valued spectrogram satisfies the following relationship:
\begin{equation}
\mathscr{F} (\mathbf{x})_{\omega, \tau} = \mathscr{F} (\mathbf{x})_{\omega, \tau-1} \, \mathrm{e}^{2 \pi j f a/L }.
\end{equation}
Therefore, the complex-valued spectrogram of a sinusoid takes the same value in each sub-band if its phase evolution is eliminated.

In order to eliminate such a phase evolution, \textit{instantaneous phase corrected STFT} (iPC-STFT) was proposed in \cite{iPCTV}:
\begin{equation}
\mathscr{F}_\mathrm{iPC} (\mathbf{x}) = \mathbf{E} \odot \mathscr{F} (\mathbf{x}),
\label{eq: iPC-STFT}
\end{equation}
where $\mathbf{E}$ is the instantaneous phase correction matrix defined by
\begin{equation}
\vspace{-2pt}
E_{\omega, \tau} = \prod_{\eta=0}^{\tau-1} \mathrm{e}^{-2 \pi j v_{\omega, \eta} a/L},
\label{eq iPCmat}
\vspace{-2pt}
\end{equation}
with $E_{\omega, 0} = 1$ for all $\omega$, and $\odot$ is the Hadamard product.
This matrix eliminates the phase evolution of a sinusoid as in Eq.~\eqref{eq: sinusoidal model}.
While real and imaginary parts of the complex-valued spectrogram calculated by the usual STFT vary owing to the phase evolution, those of iPC-STFT are constant in each sub-band thanks to the instantaneous phase correction as illustrated in Fig.~\ref{fig: illust}.
Namely, the complex-valued spectrogram of the sinusoid is smooth in each sub-band as
\begin{equation}
\mathscr{F}_{\mathrm{iPC}} (\mathbf{s})_{\omega, \tau} = \mathscr{F}_{\mathrm{iPC}} (\mathbf{s})_{\omega, \tau-1}.
\label{eq: iPC-relation}
\end{equation}
Although this equation considers a single sinusoid, the time-directional smoothness of the complex-valued spectrogram calculated by iPC-STFT is also reasonable for harmonic signals consisting of a sum of sinusoids.
In \cite{iPCTV}, based on this characteristic of iPC-STFT, a simple prior for harmonic signals was proposed, which penalizes the time-derivative of the complex-valued spectrogram calculated by iPC-STFT for enhancing harmonic components.
For more details about iPC-STFT, we refer readers to \cite{iPCTV}.

Note that the instantaneous frequency $v_{\omega,\tau}$ is not known and must be estimated in advance.
It can be estimated by the direct time-differential of phase as
\begin{equation}
\vspace{-2pt}
v_{\omega,\tau} = b\omega - \mathrm{Im} \biggl[ \frac{\tilde{\mathscr{F}} (\mathbf{x})_{\omega, \tau}}{\mathscr{F} (\mathbf{x})_{\omega, \tau}}  \biggr],
\label{eq: reasign}
\vspace{-2pt}
\end{equation}
where $\tilde{\mathscr{F}}$ is the usual STFT whose window is the time-derivative of the original window $\boldsymbol{g}$, and $\mathrm{Im}[z]$ is the imaginary part of $z$ \cite{Reasign1}.

\vspace{-2pt}
\subsection{Proposed optimization-based HPSS method}
\vspace{-2pt}

As described in the previous subsection, complex-valued spectrograms of harmonic components calculated by iPC-STFT have the distinctive structure.
Utilizing iPC-STFT, we propose a phase-aware HPSS method through the following convex optimization:
\begin{equation}
\begin{array}{cl}
    \displaystyle\min_{\mathbf{x}_\mathrm{h}, \mathbf{x}_\mathrm{p}} &\!\!\! \displaystyle\frac{1}{2}\left\| \mathbf{W} \odot D_\tau ( \mathbf{X}_\mathrm{h}) \right\|_\mathrm{Fro}^2 +　\lambda \left\| \mathbf{X}_\mathrm{p} \right\|_{2, 1} \\[8pt]
\text{s.t.} &\!\! \displaystyle \mathbf{x} = \mathbf{x}_\mathrm{h} \!+ \mathbf{x}_\mathrm{p}, \;\;  \mathbf{X}_\mathrm{h} \!= \mathscr{F}_\mathrm{iPC} (\mathbf{x}_\mathrm{h}), \;\; \mathbf{X}_\mathrm{p} \!= \mathscr{F} (\mathbf{x}_\mathrm{p})
\end{array},
\label{eq: prop}
\end{equation}
where $\mathbf{x}_\mathrm{h}$, $\mathbf{x}_\mathrm{p}$, and $\mathbf{x}$ are the time-domain signals of harmonic components, percussive components, and the audio mixture, respectively, $\lambda \! > \! 0$ is the hyperparameter which adjusts the amount of harmonic and percussive components, and $\mathbf{W} \in \mathbb{R}_+^{K \times T}$ is a weight constructed in advance.
Note that the the instantaneous phase correction matrix $\mathbf{E}$ in iPC-STFT is calculated from the the audio mixture, and thus iPC-STFT is treated as the fixed linear operator in the proposed formulation.

The first term induces the time-directional smoothness of harmonic components as an extension of Eq.~\eqref{eq: iPC-relation}.
Since power spectrogram of harmonic components is smooth in the time-direction as in Eq.~\eqref{eq: hsmooth}, the time-directional smoothness of iPC-STFT spectrogram of general harmonic components is also reasonable.
The weight $\mathbf{W}$, which adjusts the time-directional smoothness around each time-frequency bin, is given by
\begin{equation}
W_{\omega, \tau} = {\kappa}\,/\,{\max(\,\kappa,\,| \tilde{X}_\mathrm{h} |_{\omega, \tau})},
\end{equation}
where $\kappa \! > \! 0$ is a small number for adjusting the weight, and $ | \tilde{X}_\mathrm{h} |$ is the normalized amplitude of pre-estimated harmonic components which can be obtained by any existing method as \cite{HPSS2008, HPSS2010, HPSS2014KAM}.
This weight takes a small value when the amplitude of harmonic components are large, and thus harmonic components with large amplitude are not penalized so much. 
Note that this additional weight was not introduced in the previous study \cite{iPCTV}, and thus it is one of the contributions of this paper.

The second term is the $\ell_{2, 1}$-norm which induces group sparsity.
Here, the time-frame-wise sparsity is promoted by defining it as
\begin{equation}
\left\| \mathbf{X} \right\|_{2, 1} = \sum_{\tau=1}^{T}\biggl(\sum_{\omega=1}^{K} | X_{\omega, \tau}|^2 \biggr )^{\!\frac{1}{2}}.
\end{equation} 
This penalty function concentrates the energy into a few time frames and enhances impulsive components. 
Such time-frame-wise sparsity of percussive components can be seen in Fig.~\ref{fig: HP}.

\vspace{-2pt}
\subsection{Relation to the conventional methods}
\label{sec: relation}
\vspace{-2pt}

The proposed method is related to the method based on anisotropic smoothness \cite{HPSS2008} described in Section~\ref{sec: ono}.
While the first term in Eq.~\eqref{eq: ono} only considers the power spectrogram of harmonic components, that of Eq.~\eqref{eq: prop} treats both amplitude and phase through the operation in the complex domain (note that the squared amplitude of $\mathbf{X}_\mathrm{h}$ corresponds to $\mathbf{H}$).
Thus, the proposed method given by Eq.~\eqref{eq: prop} can be interpreted as a phase-aware extension of Eq.~\eqref{eq: ono}.
For the percussive components, the proposed method assumes the time-frame-wise sparsity, while Eq.~\eqref{eq: ono} considered smoothness in the frequency direction.
Considering the frequency-directional smoothness for the complex-valued spectrogram of percussive components may not be easy.
Thus, the proposed method utilizes $\ell_{2, 1}$-norm, which is insensitive to phase, instead of penalizing the frequency-derivative of the complex-valued spectrogram.

As in Eq.~\eqref{eq: prop}, the proposed method directly treats the time-domain signals, and its constraint claims that the separated components satisfy the perfect reconstruction property in the time-domain as in a recent audio source separation method \cite{TSF}.
Some of the conventional HPSS methods (e.g., anisotropic smoothness based methods \cite{HPSS2008, HPSS2014Ono} and non-negative matrix factorization based methods \cite{HPSS2014Eur, HPSS2018J}) assume additivity of power spectrograms, but it requires some statistical assumptions as discussed in \cite{ISNMF}. 
In contrast, the constraint in the proposed method (additivity in the time-domain) is always justified.
To the best of our knowledge, this is the first study applying the perfect reconstruction constraint to the separated time domain signals in HPSS.

\vspace{-2pt}
\subsection{Primal-dual splitting algorithm for proposed HPSS method}
\vspace{-2pt}
In order to solve the convex optimization problem given by Eq.~\eqref{eq: prop}, in this paper, a primal-dual splitting algorithm \cite{PDS} is adopted because it can handle some priors tied with linear operators with a constraint.
A primal-dual splitting (PDS) method \cite{PDS} is one of the convex optimization algorithms for solving the following problem:%
\footnote{
This primal-dual splitting algorithm is a simplified version chosen for easier explanation of the proposed algorithm.
We refer readers to \cite{PDS} for the general form which can handle wider range of problems.}
\begin{equation}
\min_{\mathbf{x}} \,\,\, \Theta(\mathbf{x}) + \Upsilon_1 \bigl(\mathscr{L}_1 (\mathbf{x}) \bigr) + \Upsilon_2 \bigl(\mathscr{L}_2 (\mathbf{x}) \bigr),
\vspace{-2pt}
\end{equation}
where $\Theta$ and $\Upsilon_m$ are proper lower-semicontinuous convex functions, and $\mathscr{L}_m$ is a linear operator $(m\in\{1,2\})$.
A PDS algorithm solves this problem by iterating the following procedure:
\begin{align}
\tilde{\mathbf{x}} &= \mathrm{prox}_{\mu_1 \Theta}\bigl( \mathbf{x} - \mu_{1}  \bigl(\mathscr{L}_1^{*\!} (\mathbf{y}_1^{[n]\!}) + \mathscr{L}_2^{*\!} (\mathbf{y}_2^{[n]\!})\bigr)\bigr),\!\! \\[-1pt]
\mathbf{z}_m  &= \mathbf{y}_m^{[n]} + \mathscr{L}_m^* (2\tilde{\mathbf{x}}  - \mathbf{x}^{[n]}) \quad (\forall m), \\[1pt]
    \tilde{\mathbf{y}}_m  &= \mathbf{z}_m - \mu_2 \,\mathrm{prox}_{1/\mu_2 \Upsilon_m\!} ( \mathbf{z}_m /\mu_2) \quad (\forall m), \! \\
\!\!\!\!\!\!(\mathbf{x}^{[n+1]}\!, \mathbf{y}_m^{[n+1]\!})  &= \alpha (\tilde{\mathbf{x}}, \tilde{\mathbf{y}}_m) + (1-\alpha) (\mathbf{x}^{[n]}, \mathbf{y}_m^{[n]}) \quad (\forall m), \!
\end{align}
where $\mathscr{L}_m^*$ is the adjoint operator of $\mathscr{L}_m$, $n$ is the iteration index, $\mu_1>0$, $\mu_2>0$, and $0<\alpha<2$. The important feature of this procedure is that the minimization of each function is handled separately through the proximity operator \cite{Prox}:
\begin{equation}
    \mathrm{prox}_{\rho \Psi(\mathbf{Y})} = \arg\min_{\mathbf{X}}\; \Psi(\mathbf{X}) + \frac{1}{2\rho} \left\| \mathbf{Y} - \mathbf{X} \right\|_{\mathrm{Fro}}^2.
\end{equation}

To apply this PDS algorithm to the proposed method in Eq.~\eqref{eq: prop}, it should be reformulated to the following equivalent problem:
\begin{equation}
\min_{\mathbf{x}_\mathrm{h}, \mathbf{x}_\mathrm{p}}  \,\,\, \iota_\mathbf{x}(\mathbf{x}_\mathrm{h}, \mathbf{x}_\mathrm{p}) + \frac{1}{2}\left\| \mathscr{L}_\mathrm{h} (\mathbf{x}_\mathrm{h}) \right\|_\mathrm{Fro}^2 + \lambda
\left\| \mathscr{F} (\mathbf{x}_\mathrm{p}) \right\|_{2, 1},
\label{eq: reprop}
\vspace{-2pt}
\end{equation}
where $\mathscr{L}_\mathrm{h}(\cdot)= \mathbf{W} \odot D_\tau \bigl( \mathscr{F}_\mathrm{iPC} (\cdot) \bigr)$,
and $\iota_\mathbf{x} (\cdot, \cdot)$ is the indicator function of the perfect reconstruction constraint given by
\begin{equation}
\iota_\mathbf{x}(\mathbf{x}_\mathrm{h}, \mathbf{x}_\mathrm{p}) = \left\{
\begin{array}{cl}
0&(\mathbf{x} = \mathbf{x}_\mathrm{h} + \mathbf{x}_\mathrm{p}) \\
\infty &(\text{otherwise})
\end{array}
\right..
\end{equation}
Applying the PDS algorithm to the reformulated problem in Eq.~\eqref{eq: reprop} yields Algorithm~\ref{alg: reprop}, where the choice of $\mu_1$ and $\mu_2$ can be automated as in \cite{PDS}, and the proximity operators involved in the algorithm can be analytically calculated as follows \cite{Prox, MixNorm}:
\begin{align}
P_\mathbf{x} (\mathbf{x}_\mathrm{h}, \mathbf{x}_\mathrm{p}) &=  (\mathbf{x}_\mathrm{h}, \mathbf{x}_\mathrm{p}) + (\mathbf{x} - \mathbf{x}_\mathrm{h} - \mathbf{x}_\mathrm{p})/2, \\[3pt]
\mathrm{prox}_{\rho\| \cdot \|_\mathrm{Fro}^2}(\mathbf{X}) &=  \mathbf{X}/(1+\rho), \\
( \mathrm{prox}_{\rho \| \cdot \|_{2, 1}}(\mathbf{X}) )_\tau &= \left(1-\rho/\| \mathbf{X}_\tau \|_2 \right)_+ \, \mathbf{X}_\tau,
\end{align}
where $\mathbf{X}_\tau$ is the $K$-dimensional vector at the $\tau$th time frame.
We stress that this algorithm does not require the inverse of linear operators, $\mathscr{L}_\mathrm{h}$ and $\mathscr{F}$, but only require applying them and their adjoint, and thus we can avoid the huge computation for calculating their inverse.

\begin{algorithm}[t!]                      
\caption{Proposed HPSS algorithm solving Eq.~\eqref{eq: reprop}}         
\label{alg: reprop}
\begin{algorithmic}
\STATE \textbf{Input}: $\mathbf{x}$, $\mathbf{x}_\mathrm{h}^{[0]}$, $\mathbf{x}_\mathrm{p}^{[0]}$, $\mathbf{Y}_\mathrm{h}^{[0]}$, $\mathbf{Y}_\mathrm{p}^{[0]}$, $\lambda$, $\mu_1$, $\mu_2$, $\alpha$
\STATE \textbf{Output}: $\mathbf{x}_\mathrm{h}^{[n+1]}$, $\mathbf{x}_\mathrm{p}^{[n+1]}$
\FOR {$n = 1, 2, \ldots$}
\STATE $(\tilde{\mathbf{x}}_\mathrm{h}, \tilde{\mathbf{x}}_\mathrm{p}) = P_\mathbf{x} \bigl(\mathbf{x}_\mathrm{h}^{[n]} - \mu_1 \mathscr{L}_\mathrm{h}^*(\mathbf{Y}_\mathrm{h}^{[n]}),\; \mathbf{x}_\mathrm{p}^{[n]} - \mu_1\mathscr{F}^*(\mathbf{Y}_\mathrm{p}^{[n]}) \bigr)$
\STATE $\mathbf{z}_\mathrm{h} = \mathbf{y}_\mathrm{h}^{[n]} + \mathscr{L}_\mathrm{h}(2\tilde{\mathbf{x}}_\mathrm{h} - \mathbf{x}_\mathrm{h}^{[n]})$
\STATE $\mathbf{z}_\mathrm{p} = \mathbf{y}_\mathrm{p}^{[n]} + \mathscr{F}(2\tilde{\mathbf{x}}_\mathrm{p} - \mathbf{x}_\mathrm{p}^{[n]})$
\STATE $\tilde{\mathbf{y}}_\mathrm{h} = \mathbf{z}_\mathrm{h} -\mu_2 \,\mathrm{prox}_{(1/\mu_2) \|\cdot\|_{\text{Fro}}^2} (\mathbf{z}_\mathrm{h}/\mu_2)$
\STATE $\tilde{\mathbf{y}}_\mathrm{p} = \mathbf{z}_\mathrm{p} -\lambda \mu_2 \,\mathrm{prox}_{(1/\lambda \mu_2) \|\cdot\|_{2, 1}} (\mathbf{z}_\mathrm{p}/\lambda \mu_2)$
\STATE $(\mathbf{x}^{[n+1]}_{\mathrm{h}, \mathrm{p}}, \mathbf{y}_{\mathrm{h}, \mathrm{p}}^{[n+1]}) = \alpha (\tilde{\mathbf{x}}_{\mathrm{h}, \mathrm{p}}, \tilde{\mathbf{y}}_{\mathrm{h},\mathrm{p}}) + (1-\alpha) (\mathbf{x}^{[n]}_{\mathrm{h}, \mathrm{p}}, \mathbf{y}_{\mathrm{h},\mathrm{p}}^{[n]})$
\ENDFOR
\end{algorithmic}
\end{algorithm}

\vspace{-2pt}
\section{Numerical experiment}
\vspace{-2pt}
The proposed method was applied to separations of $10$ audio tracks%
\footnote{It can be downloaded from \url{https://www.idmt.fraunhofer.de/en/business_units/m2d/smt/phase_based_harmonic_percussive_separation.html}}
which was presented in the previous study \cite{HPSS2014Phase}.
The sampling rate was $44100$ Hz, and STFT was calculated by the canonical tight window of the Hann window of $4096$ samples with $1024$ sample shifting.
The proposed method was compared with four conventional methods: Ono's \cite{HPSS2008}, MF \cite{HPSS2010}, KAM \cite{HPSS2014KAM}, and PM \cite{HPSS2014Phase}.
In each method, the hyperparameters were set to suggested values in each original paper.
Separation performance was evaluated by the average of BSS Eval measures: SDR, signal-to-interference ratio (SIR), signal-to-artifacts ratio (SAR) \cite{BSS}.

For the proposed method, the amplitude spectrogram of harmonic components should be estimated in advance for calculating the weight $\mathbf{W}$.
Here we utilized MF as a simple and fast HPSS method for prior estimation of the harmonic components.
Although MF does not treat phase, it is modified through the proposed method.
For calculating iPC-STFT, the instantaneous frequency of the harmonic components is also required.
Since its oracle information is not available, we calculated it from the audio mixture (Prop-mix).
In order to evaluate the potential of the proposed method, iPC-STFT with the instantaneous frequency calculated from the oracle harmonic components was also compared (Prop-ora).
In both cases, the instantaneous frequency was calculated by Eq.~\eqref{eq: reasign}.
The hyperparameters, $\lambda$ and $\kappa$, were experimentally determined to be $0.5$ and $0.001$.
Algorithm~\ref{alg: reprop} was iterated $100$ times with $\mu_1 = 1$, $\mu_2 = 0.25$, and $\alpha = 0.5$.
While an arbitrary choice is allowable, MF was utilized for estimating the initial value.

{\renewcommand\arraystretch{1.1}
\begin{table}
\caption{Mean scores over $10$ audio tracks. The average (Ave.) of the harmonic (Har.) and percussive (Per.) components are also presented. Bold font indicates the highest (excluding Prop-ora).}
\vspace{2pt}
\centering
\footnotesize
\scalebox{0.975}{
	\begin{tabular}{c| c | c c c c | c | c}
\hline
	   &        & \!\!Ono's\! \cite{HPSS2008}\!\! & \!\!\!MF\! \cite{HPSS2010}\!\!\! & \!\!\!KAM\! \cite{HPSS2014KAM}\!\!\! & \!\!PM\! \cite{HPSS2014Phase}\!\! &\!\!Prop-mix\!\!&\!\!Prop-ora\!\!\\
\hline
	\multirow{3}{*}{\!\!\!Har.\!\!\!} & \!\!SDR\!\! &$5.8$ &$8.6$&$4.9$&$-8.6$&$\mathbf{9.3}$&$10.3$ \\ 
     & \!\!SIR\!\! & $11.2$&$15.1$&$\!\!\mathbf{23.1}\!\!$&$6.1$&$12.3$&$13.8$ \\
     & \!\!SAR\!\! &$7.6$ &$10.2$&$5.1$&$-7.5$&$\mathbf{15.4}$&$15.4$ \\
	\hline
	\multirow{3}{*}{\!\!\!Per.\!\!\!} & \!\!SDR\!\! &$-8.1$&$-4.2$&$-4.7$&$-12.1$&$\mathbf{-3.8}$&$-2.7$ \\ 
    & \!\!SIR\!\! &$-2.8$ &$-1.3$&$-2.3$&$-3.2$ &$\mathbf{1.7}$ &$2.8$ \\
    & \!\!SAR\!\! &$-1.9$&$3.5$&$\mathbf{4.2}$&$-6.7$   &$1.6$ &$2.6$ \\
	\hline\hline
	\multirow{3}{*}{\!\!\!Ave.\!\!\!}& \!\!SDR\!\! &$-1.1$&$2.0$&$0.1$&$-10.4$&$\mathbf{2.8}$&$3.8$  \\ 
    & \!\!SIR\!\!  &$4.2$ &$6.9$&$\!\!\mathbf{10.4}\!\!$&$1.5$&$7.0$&$8.3$ \\
    & \!\!SAR\!\! &$2.8$ &$6.9$& $4.6$&$-7.1$&$\mathbf{8.5}$&$9.0$ \\
	\hline
	\end{tabular}
}
\label{tab: result}
\end{table}
}

\vspace{-2pt}
\subsection{Results}
\vspace{-2pt}
The experimental results are summarized in Table~\ref{tab: result}.
MF outperformed other conventional methods in terms of SDR, which served as the initial value of the proposed method.
Since the time-frequency mask was estimated solely from the phase information, PM resulted in the lowest performance.
In contrast, the proposed method, which simultaneously treats both amplitude and phase, outperformed conventional methods in terms of SDR.
We observed that KAM reduced much components, which significantly improved SIR of harmonic components but induced low SDR and SAR.
As a result of taking phase into account, the proposed method achieved higher scores than Ono's which is the non phase-aware version of the proposed method as discussed in Section~\ref{sec: relation}.

Comparing Prop-ora with Prop-mix, we confirmed that the accurate estimation of the instantaneous frequency of harmonic components can improve the performance of the proposed method.
This is because the first term in Eq.~\eqref{eq: prop} takes a smaller value for harmonic components and penalizes percussive components more by utilizing the appropriate instantaneous frequency.
The instantaneous frequency estimation given by Eq.~\eqref{eq: reasign} is one of the simplest methods.
More specific methods for the accurate estimation of the instantaneous frequency of harmonic components should be considered, which is a future work.

\vspace{-2pt}
\section{Conclusion}
\vspace{-2pt}

In this paper, a phase-aware HPSS method through convex optimization was proposed.
Based on two HPSS approaches (anisotropic smoothness and sinusoidal model), the proposed method assumes the smoothness of the complex-valued spectrogram of harmonic components calculated by iPC-STFT in the time direction.
On the other hand, the time-frame-wise sparsity of percussive spectrograms was considered as a phase insensitive prior.
Furthermore, the proposed method considers the perfect reconstruction constraint in the time domain instead of power spectrograms.
Through the experiment, the effectiveness of the proposed method was validated in terms of SDR.
The experimental results indicated the accurate estimation of the instantaneous frequency of harmonic components can improve the performance of the proposed method, which is included in our future works.
\clearpage
\bibliographystyle{IEEEbib}

\end{document}